# Electrical Probing of Dark Excitons through Microwave Permittivity

Alex Stram[1†], Zhida Liu[1†], Ziheng Zhang[2†], Yangchen He[3], Xuejian Ma[1], Kyoung Pyo Lee[1], Lisa Frammolino[1], Fuxiang Chen[1], Robert J. Boyd[3], Sirapas Tangton[1], Anand Swain[5], Kan Yao[6], Kenji Watanabe[7], Takashi Taniguchi[8], Yuebing Zheng[5,6], Chih-Kang Shih[1], Daniel Rhodes[3,4], Li Yang[2*], Xiaoqin Li[1*], Keji Lai[1*]

[1]*Department of Physics, University of Texas at Austin, Austin TX 78712, USA*
[2]*Department of Physics, Washington University in St Louis, St Louis, MO 63130, USA*
[3]*Department of Materials Science and Engineering, University of Wisconsin-Madison, Madison, WI 53706, USA*
[4]*Department of Physics, University of Wisconsin-Madison, Madison, WI 53706, USA*
[5]*Texas Materials Institute, The University of Texas at Austin, Austin 78712, USA*
[6]*Walker Department of Mechanical Engineering, Materials Science and Engineering Program, University of Texas at Austin, Austin TX 78712, USA*
[7]*Research Center for Electronic and Optical Materials, National Institute for Materials Science, 1-1 Namiki, Tsukuba 305-0044, Japan*
[8]*Research Center for Materials Nanoarchitectonics, National Institute for Materials Science, 1-1 Namiki, Tsukuba 305-0044, Japan*

[†] These authors contributed equally to this work
[*] Corresponding authors

## Abstract

Excitons in atomically thin semiconductors are almost always probed through their optical signatures, because the short lifetimes of these transient quasiparticles are generally assumed to preclude electrical detection. Here we show that photoexcited excitons in monolayer tungsten disulfide produce a large, optically tunable permittivity at gigahertz frequencies, and that the effect provides a contact-free electrical route to imaging dark excitons at the nanoscale. Using laser-illuminated microwave impedance microscopy, we find that high-purity encapsulated flakes exhibit a purely dielectric response resonant with the exciton spectrum, whereas defect-rich samples are governed by conventional photoconductivity. Spatial mapping of diffusion and sublinear power dependence identify long-lived dark excitons as the dominant contributors, and first-principles modelling of exciton polarizability reproduces the measured susceptibility. Our results establish excitons as optically tunable dielectric elements and introduce microwave microscopy as a direct electrical probe of dark-exciton transport with sub-100 nm resolution.

Excitons are central to the optical and optoelectronic properties of semiconductors. In atomically thin materials, reduced dielectric screening and quantum confinement produce tightly bound excitons, which remain stable at room temperatures and may contribute to energy transport in optoelectronic devices[1,2]. These neutral quasiparticles do not undergo drift motion in an electric field, but their internal electron-hole coordinates can be displaced, giving rise to an induced dipole moment. In this sense, an ensemble of excitons should contribute to the dielectric response of a material, analogous to the way microscopic polarizability gives rise to macroscopic permittivity in molecular or atomic media through the Clausius-Mossotti relation[3]. However, optically generated excitons are not permanent constituents of the solid. They are instead transient, non-equilibrium quasiparticles whose lifetimes can range from picoseconds to nanoseconds for bright excitons[4] and orders of magnitude longer for dark species[5,6]. Whether such an exciton population can produce a quasistatic susceptibility significant enough for electronic devices remains an open question. Moreover, for dark excitons that are invisible to conventional spectroscopy, a low-frequency permittivity would allow them to be probed electrically, instead of relying on indirect optical probes following conversion to bright states.

The dielectric response of excitons is particularly relevant for two-dimensional (2D) transition metal dichalcogenides (TMDs), where excitons feature large binding energies[7], strong oscillator strengths[8], spin-valley locking[9], and a rich hierarchy of bright, dark, charged and excited states[2,10]. At optical frequencies, excitonic polarizability is usually probed through absorption[11] and pump-probe spectroscopy and microscopy[12]. In the opposite DC limit, electric fields are often used to dissociate excitons or collect photo-generated carriers, so the measured response is dominated by photocurrent[13]. Between these limits lies the microwave regime, where the electric field varies slowly compared with optical cycles. The response of excitons in this intermediate frequency range is much less understood, despite its potential relevance to optically tunable capacitors, microwave photonics, and nanoscale electronic sensing.

Here we address these questions using monolayer tungsten disulfide ($WS_2$) as a model system and laser-illuminated microwave impedance microscopy (iMIM)[14] to probe the gigahertz (GHz) response of photoexcited excitons. The microwave probe senses local changes in the complex tip-sample admittance, allowing dielectric and conductive responses to be separated without electrical contacts. We find that defect-rich $WS_2$ is dominated by conventional photoconductivity, whereas high-purity hBN-encapsulated monolayers exhibit a purely capacitive response resonant with the exciton absorption spectrum. The extracted in-plane dielectric constant reaches values more than twice that of the intrinsic monolayer, indicating that a steady-state ensemble of excitons can produce a giant effective permittivity at GHz frequencies. Spatial diffusion mapping and laser-power-dependent measurements point to long-lived dark excitons as the main contributors to this dielectric screening. Supported by theoretical modelling of exciton polarizability and local-field effects, our results establish photoexcited excitons as optically tunable dielectric elements for microwave electronics. Unlike a recent microwave measurement[15] that infers excitons from photoconductivity and Auger-assisted tunneling, the clean-limit response reported here is purely dielectric, providing a contact-free electrical readout of dark excitons that does not require the conversion to bright states or free carriers.

The schematic of our experiments is illustrated in Fig. 1A, with further details provided in Supplementary Material S1. We use a supercontinuum laser as the excitation source, the output of which is directed through a mechanical grating monochromator for wavelength tuning. To ensure a constant incident power across the spectral range, the beam is regulated by an electrically controlled neutral density (ND) filter. The laser is focused onto the sample through the transparent substrate using an objective lens mounted on an XY-stage. From the opposite side, the iMIM tip is mounted on an XYZ scanner to approach the sample surface. Both the sample and the tip are placed inside a flow cryostat with a quartz window for optical access.

A central finding of this work is that resonantly generated excitons can produce a remarkably large dielectric response at GHz frequencies. As summarized in Fig. 1B, at temperature $T$ = 80 K and under an average laser power $P$ = 0.38 μW, an encapsulated monolayer $WS_2$ device exhibits a strongly photon-energy-dependent effective dielectric constant at 1 GHz. The spectrum displays pronounced peaks at $WS_2$ exciton resonances, directly linking the enhanced dielectric response to the polarizability of photoexcited excitons. The measured in-plane relative permittivity or dielectric constant $\kappa_{\parallel} = \varepsilon_{\parallel}/\varepsilon_0$, where $\varepsilon_0$ is the vacuum permittivity, can be as large as ~ 30, which is more than double of that in monolayer $WS_2$ with $\kappa_{\parallel}$ = 14 [Ref. 16]. The uncertainty of conversion from the iMIM data to dielectric constant is around 20 ~ 30% (Supplementary Material S2), which does not change our conclusion here. Such a large, resonantly enhanced GHz permittivity suggests that excitons can substantially modify the dielectric landscape of surrounding materials in radio-frequency (RF) electronic devices.

Unlike photocurrent spectroscopy[17,18], iMIM enables contact-free electrical detection under optical excitation. Given that the tip diameter (~ 100 nm) is roughly six orders of magnitude smaller than the 1 GHz electromagnetic wavelength, the tip-sample interaction can be accurately modeled by lumped elements[19]. During illumination, changes in the sample electrical properties (permittivity $\varepsilon$ or conductivity $\sigma$) induce variations in the real (Re) and imaginary (Im) parts of the tip admittance. The iMIM signals can be understood by finite-element analysis (FEA)[14,19]. As shown in Fig. 2A, when $\sigma$ increases from zero but remains small (~ 1 S/m), signals in both channels rise monotonically, with iMIM-Re being dominant. Conversely, an increase in $\varepsilon$ from its background value produces a response exclusively in the Im channel, while the Re-part signal remains zero. Fig. 2B shows the simulated iMIM signals as a function of the effective in-plane dielectric constant $\kappa_{\parallel} = \varepsilon_{\parallel}/\varepsilon_0$. Here we ignore the out-of-plane polarizability of excitons since it is negligibly small compared to the in-plane counterpart (see theoretical analysis below). Quantitative comparison between the simulated and measured iMIM data can be obtained through calibration of the iMIM electronics[20], as detailed in Supplementary Material S2.

We first present results that can be fully accounted for by pure photoconductivity response. As shown in Fig. 2C, for CVD-grown $WS_2$ with $Al_2O_3$ coating[14], hereafter labeled as Sample $a$, a clear onset of iMIM signal (taken at $T = 293$ K and $P = 24$ μW) is observed at $h\nu \sim 2.2$ eV in the spectrum, corresponding to the electronic gap of monolayer $WS_2$. Moreover, since the iMIM-Re signal is larger than the iMIM-Im counterpart, the data indicates that the response is dominated by $\sigma$ in $WS_2$ (Supplementary Material S2). The same phenomena have been reported in previous iMIM works on various semiconductors[14,15,21-24]. The underlying physical picture is illustrated in Fig. 2D. Upon photon absorption, electrons and holes are rapidly generated and thermalized to the band extrema. In the presence of a high defect density (~ $10^{12}$ cm$^{-2}$ in Sample $a$, as previously reported in Ref. [14]), holes in $WS_2$ are trapped by band-tail states, leaving mobile electrons in the conduction band that are responsible for the observed photoconductivity.

Strikingly, the iMIM data on ultra-clean monolayer $WS_2$ flakes exfoliated from flux-grown crystals and embedded between hBN layers, hereafter labeled as Sample $b$, exhibits completely different characteristics. The density of charged defects in Sample $b$ is well below $10^{10}$ cm$^{-2}$, as confirmed by conductive atomic-force microscopy (C-AFM) measurements[25] (Supplementary Material S3). First and foremost, as shown in Fig. 2E, the iMIM-Re signal remains zero for the entire spectrum up to $h\nu$ ~ 2.6 eV, which is much higher than the electronic gap. Secondly, the onset of iMIM-Im signal shifts to a lower photon energy $h\nu$ ~ 1.9 eV, i.e., the optical gap of $WS_2$, than that of Sample $a$. Thirdly, the iMIM-Im signal displays pronounced peaks when the photon energy matches the A- (2.06 eV) and B-exciton (2.45 eV) resonances. The shoulder-like features at 2.03 eV and 2.23 eV are presumably associated with the trion and 2s states of A-exciton, respectively. Finally, using FEA simulations, we can convert the iMIM data to the effective dielectric constant of monolayer $WS_2$ under illumination, as presented before in Fig. 1B. This purely dielectric response is distinct from that of a recent iMIM study[15], in which excitons were detected through photoconductivity changes and Auger-assisted tunneling.

The absence of iMIM-Re signals in Fig. 2E suggests that photoconductivity from mobile carriers is negligible in Sample $b$. When the defect density is sufficiently low, the majority of photo-generated electron-hole pairs thermalize to the band edges and form neutral excitons, as illustrated in Fig. 2F. Under the laser excitation, a steady-state population of A-excitons is maintained, whose collective response manifests as an effective permittivity. Consequently, the iMIM-Im data in Fig. 2E primarily reflects the steady-state exciton density and closely resembles the characteristic $WS_2$ absorption spectrum reported in the literature[11,26]. Notably, residual photoconductivity signals are observed in defective regions of Sample $b$, such as edges, wrinkles, or bubbles (Supplementary Material S4). In these areas, the underlying physics lies between the defect-rich and high-purity limits illustrated in Fig. 2D and Fig. 2F, respectively. In addition, we also measured the iMIM spectra on an uncapped monolayer $WS_2$ sample (Supplementary Material S5), where strong peak-like features are observed in both Im and Re channels at the exciton resonances. We attribute this to the lack of an hBN spacer between the tip and $WS_2$. The GHz electric field is strong enough to dissociate an appreciable portion of excitons, resulting in mixed conductivity and permittivity responses.

Further insights on the exciton permittivity can be obtained from iMIM experiments under various conditions. Fig. 3A shows the spatial iMIM-Im mapping when the tip is parked on a flat region of Sample $b$ and the laser beam is raster scanned. In contrast to the localized laser profile captured by a CCD camera (inset of Fig. 3A), the iMIM-Im data displays significant spatial broadening with a diffusion length $\lambda$ ~ 12 μm, as plotted in Fig. 3B. Details of the diffusion analysis[14] are included in Supplementary Material S6. In W-based monolayer TMDs, optical absorption occurs between the top valence band and the second conduction band due to strong spin-orbit coupling and spin selection rules. These bright excitons can quickly convert to dark excitons consisting of electrons at the lowest conduction band [27-29]. The result is the predominance of spin-forbidden dark excitons with a long lifetime and a large diffusion length[30] on the order of 10 μm, which is consistent with our observation.

The $T$-dependent iMIM-Im spectra of Sample $b$ are plotted in Fig. 3C. As shown in Supplementary Material S7, the data can be accounted for by four ($A_T$, $A_{1S}$, $A_{2S}$, and $B_{1S}$) Lorentzian peaks and a slow-varying background. The optical differential reflectance curves of Sample $b$ (Supplementary Material S8) are also overlaid in Fig. 3C. The correspondence between the most prominent iMIM peak and the optical signature of $A_{1S}$ resonance at each temperature confirms that the microwave signal indeed originates from excitons. Finally, Fig. 3D shows the power-dependent iMIM spectra when the photon energy matches the A- and B-exciton resonances. The power dependence is strongly sub-linear even for $P < 1$ μW and rapidly saturates as the laser power exceeds ~ 3 μW. At low excitation power, the steady-state exciton density $n_X$ is small (on the order of $10^{11}$ cm$^{-2}$) and proportional to the photon flux[31] (Supplementary Material S9). Owing to the reduced screening in monolayer TMDs, the bimolecular exciton-exciton annihilation (EEA)[32] becomes significant at an intermediate $n_X \sim 10^{12}$ cm$^{-2}$, beyond which the exciton density no longer increases linearly with the laser power. Since the effective dielectric constant of excitons roughly scales with $n_X$ (see theoretical analysis below), the observed sub-linear power dependence is most likely associated with the EEA process. On the other hand, when $n_X$ is as high as $10^{13} \sim 10^{14}$ cm$^{-2}$, the system is expected to transition from an insulating gas of neutral excitons to a metallic electron-hole plasma[33]. The iMIM-Re signal remains zero in Fig. 3C, indicating that the many-body Mott transition of excitons has not been reached under our excitation conditions.

To interpret the iMIM results, we theoretically investigate the microscopic polarizability and macroscopic permittivity of excitons in TMDs. We use an effective Keldysh electron-hole Hamiltonian to calculate the exciton wavefunctions[34-37]. The screening length $r_0$ in the Keldysh potential is determined by fitting the triplet dark-exciton binding energy obtained from our GW-BSE calculation, so that the effective model is consistent with the dark-exciton response probed experimentally. Figs. 4A and 4B show the calculated exciton wavefunctions in monolayer $WS_2$ embedded in hBN under in-plane external fields of $E_{\text{ext}}$ = 0 and 0.1 V/nm, respectively. Note that the typical GHz electric field in the experiment is on the order of 0.01 V/nm, whereas the large $E_{\text{ext}}$ used in Fig. 4B is to magnify the field distortion for illustration. In the weak-field regime, the exciton polarizability α can be extracted from the quadratic Stark shift $\Delta = -\frac{1}{2}\alpha E_{\text{ext}}^2$. Using the exciton wavefunctions obtained from the effective Keldysh electron-hole Hamiltonian, the exciton polarizability can be evaluated from second-order perturbation theory[38-40]. For an in-plane electric field, the perturbation term is $H' = e\vec{\mathbf{E}}_{\mathbf{ext}} \cdot \vec{\mathbf{r}}$, where $\vec{\mathbf{r}}$ is the electron-hole relative coordinate. The angular part of the perturbation term only couples the 1$s$ ground state to the $p$ exciton states, resulting in an in-plane polarizability:

$$\alpha_{\parallel} = e^2 \sum_{np} \frac{\left|\langle \mathrm{R}_{np}|r|\mathrm{R}_{1s}\rangle\right|^2}{E_{np}^{(0)} - E_{1s}^{(0)}} \tag{1}$$

, where $R_{np}$ and $R_{1s}$ denote the radial parts of the $p$ exciton states and the 1$s$ ground state, respectively (Supplementary Material S10). The out-of-plane polarizability $\alpha_{\perp}$ is estimated by a finite quantum-well model for the vertical confinement of the electron and hole[38,41,42].

Fig. 4C shows the calculated $\alpha_{\parallel}$ and $\alpha_{\perp}$ as a function of the effective isotropic dielectric constant of the substrate ($\kappa \sim 5$ for hBN[15]). Because of the atomically thin nature of monolayer $WS_2$, $\alpha_{\perp}$ is two orders of magnitude smaller than $\alpha_{\parallel}$. We will therefore ignore the out-of-plane dielectric response in further analysis. The information of exciton polarizability allows us to evaluate the excitonic contribution to the quasistatic dielectric constant, i.e., the exciton susceptibility, through a Clausius-Mossotti-like analysis[43] (Supplementary Material S10). For micron-sized 2D materials, the depolarization field generated by the boundary charge is negligible in the bulk region such that the macroscopic electric field $E_{\text{mac}} \approx E_{\text{ext}}$. The local microscopic field $E_{\text{loc}}$ felt by a single exciton, on the other hand, is the sum of $E_{\text{mac}}$ and the interaction field $E_{\text{i}}$ from surrounding excitonic dipoles, i.e., $E_{\text{loc}} = E_{\text{mac}} + E_{\text{i}}$. If $E_{\text{i}}$ is neglected, one obtains the bare exciton susceptibility $\chi_{\text{bare}} = \frac{n_X p}{\varepsilon_0 E_{\text{loc}} d} = \frac{\alpha n_X}{\varepsilon_0 d}$, where $p$ is the induced dipole moment of one exciton and $d$ is the thickness of monolayer $WS_2$. To include the interaction field from surrounding excitons, we follow the local-field correction model[43] and write the interaction field under a dielectric environment of $\kappa$ as:

$$E_{\text{i}} = \frac{1}{\kappa} \frac{\alpha C_0}{4\pi\varepsilon_0 a^3} E_{\text{loc}} \tag{2}$$

, where $C_0/4\pi = 0.3538$ is a dimensionless local-field factor and $a \sim n_N^{-1/2}$ is the average inter-exciton distance. The corrected exciton susceptibility can then be evaluated as:

$$\chi_{\text{corr}} = \frac{\chi_{\text{bare}}}{1 - \frac{0.3538}{\kappa}\sqrt{n_X}\, d\chi_{\text{bare}}} \tag{3}$$

Both $\chi_{\text{bare}}$ and $\chi_{\text{corr}}$ of hBN-encapsulated monolayer $WS_2$ are plotted in Fig. 4D. For the relevant exciton density $n_X \sim 10^{12}$ cm$^{-2}$ in our experiment, the local field correction only slightly modifies the susceptibility. More importantly, the calculated exciton susceptibility is on the order of $10^1$, which is in good agreement with the values (Fig. 1B) extracted from the iMIM data.

We conclude by discussing several implications of this work. First, our findings reveal that the effective macroscopic permittivity due to a steady-state ensemble of excitons can exceed the intrinsic permittivity of the host material. Such a giant GHz dielectric response of the 2D semiconductor can also be tuned by the wavelength and power of the excitation laser, bridging the gap between exciton physics and RF electronics. Secondly, while substantial progress has been made in visualizing exciton transport[44], the spatial resolution of optical microscopy is usually limited by diffraction. In addition, since dark excitons cannot be directly probed optically, a dark-to-bright exciton conversion process [30] is required, which complicates the data interpretation. Our results show that dark excitons can be detected through their electrical polarizability rather than through indirect optical signatures, providing a new approach to investigate exciton transport with sub-100nm spatial resolution. Thirdly, while this initial study focuses on the intralayer excitons in monolayer TMDs, the methodology can be easily extended to van der Waals heterostructures. For instance, nanoscale microwave imaging of excitons may shed light on interlayer excitons with out-of-plane dipole moments[45], the exciton localization by moiré superlattices[46], and the crossover from an exciton Bose-Einstein condensate to an excitonic insulator[47]. Finally, the formalism of Stark-effect approach and local-field corrections is generally applicable to various excitonic systems. Our theoretical work thus establishes the foundation to relate microscopic polarizability and macroscopic permittivity of different exciton species. In all, our work offers a new paradigm for developing optoelectronic devices with optically tunable dielectric functions in the microwave regime, as well as a new approach to explore optical physics with electrical imaging.

## Materials and Methods

### Sample preparation

Two types of monolayer $WS_2$ samples were studied in this work. The control samples (Sample $a$) were grown directly on double-side-polished sapphire substrates by chemical vapor deposition. Detailed growth parameters and conditions can be found in Ref. [14]. A thin dielectric layer (4 nm amorphous $Al_2O_3$) was deposited on the sample surface before the iMIM experiment. For the main samples (Samples $b1 - b3$), monolayer flakes were exfoliated from bulk crystals onto $SiO_2$/Si chips using the standard scotch-tape method. They were then picked up by polyethylene terephthalate (PET) polymer stamps and encapsulated by thin hBN layers before being transferred to double-side-polished sapphire substrates. The samples were annealed in vacuum and cleaned by contact-mode AFM scans before the iMIM experiment. We also measured an unencapsulated monolayer $WS_2$ (Sample $b4$) for comparison.

### iMIM measurement

The microwave imaging experiment was performed in a customized variable-temperature chamber (ST-500, Janis Research Co.). The PtIr probe tip (12PtIr400A, Rocky Mountain Nanotechnology LLC) was mounted on the low-temperature positioners and scanners (AttoCube Systems AG). We used a white supercontinuum laser (SuperK Fianium FIU15, NKT Photonics) as the light source. The laser output was delivered to a mechanical grating monochromator (Model 9055, Sciencetech Inc.) for wavelength sweeping. The beam was then sent through an electrically controlled neutral density filter to maintain constant laser power at each wavelength. Finally, the laser was focused on the sample surface by an objective lens attached to an XY piezo stage.

### FEA simulation

The commercial software COMSOL Multiphysics 5.4 was employed to perform finite-element analysis (FEA) and simulate the real and imaginary parts of the admittance for the specific tip-sample geometry. Details of the thickness and dielectric constant of each layer in the simulation are included Supplementary Material S2.

### Computational methods

First-principles DFT and GW-BSE calculations were performed to obtain the electronic and excitonic properties of monolayer $WS_2$. An effective Keldysh electron-hole model was then constructed by fitting the triplet exciton binding energies obtained from the GW-BSE calculations.

Based on the exciton wave functions from this effective model, the exciton in-plane polarizability was extracted using the second-order perturbation theory. The out-of-plane polarizability was estimated using a finite quantum-well model.

## Acknowledgements

The iMIM study by A.S. was primarily supported by the NSF under grant DMR-2426989 and the Welch Foundation under grant No. F-1814. K.L. acknowledges support from United States Army Research Office via grant W911NF-25-1-0232 for equipment and data analysis. The work performed by Z.L. and the collaboration between C.-K.S. and X.L. are facilitated by the National Science Foundation through the MRSEC program under Cooperative Agreement No. DMR-2308817. X.L. gratefully acknowledges support from the Office of Naval Research via grant N000142512069, Welch Foundation Chair F-0014 for sample preparation, and Air Force Office of Scientific Research under MURI grant FA9550-25-1-0288 for exciton transport studies. The collaboration between UT-Austin (K.L. and X.L.) and Washington University (L.Y.) was enabled by NSF Designing Materials to Revolutionize and Engineer our Future (DMREF) program via grants DMR-2118806 and DMR-2118779. Y.H. acknowledges support from NSF MRSEC DMR-2309000. D.R. acknowledges support from Office of Naval Research via grant N000142512276. Y.Z. acknowledges support from National Science Foundation PFI-2140985. K.W. and T.T. acknowledge support from the CREST (JPMJCR24A5), JST and World Premier International Research Center Initiative (WPI), MEXT, Japan. Research by R.J.B. was sponsored by the Army Research Laboratory and was accomplished under Cooperative Agreement Number W911NF-24-2-0054. The views and conclusions contained in this document are those of the authors and should not be interpreted as representing the official policies, either expressed or implied, of the Army Research Laboratory or the U.S. Government.

## Author contributions

X.L. and K.L. conceived the experiment. A.S. fabricated the samples, set up the experiment with assitance from Z.L. and X.M., carried out the iMIM measurement, and performed data analysis. K.P.L. performed the optical reflectance measurement. Y.H., R.J.B., and D.R. synthesized and characterized the bulk $WS_2$ crystals. L.F., F.C., and C.-K.S. synthesized the CVD $WS_2$ crystals. K.W. and T.T. synthesized the hBN crystals. Z.Z. and L.Y. proposed the theoretical model and performed the calculations. A.S. and K.L. wrote the first draft of the manuscript. All authors contributed to discussions.

## Competing interests

The authors declare no competing interests.

## Data availability

The online version contains all raw data and supplementary materials.

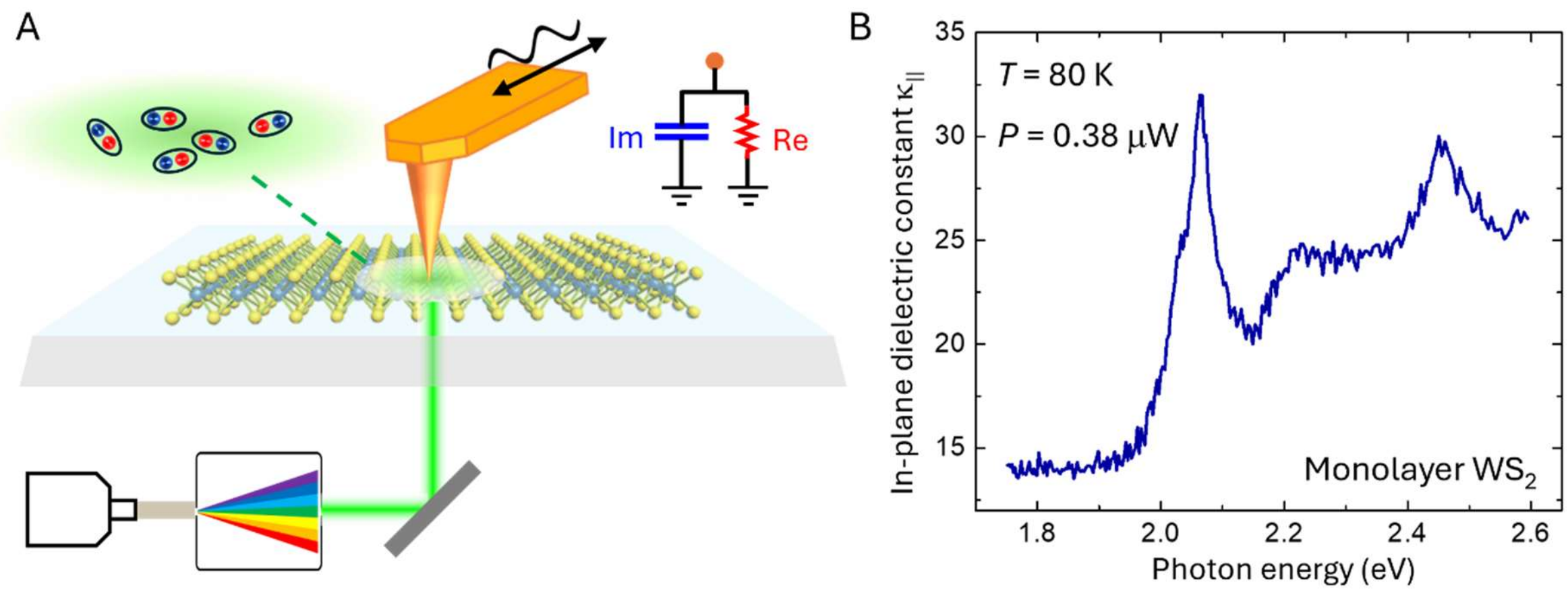

**Fig. 1. iMIM configuration and GHz dielectric constant of excitons in monolayer $WS_2$.** (**A**) Schematic of the iMIM setup. The monochromatized laser beam generates a steady state of excitons, whose dielectric response is detected by the iMIM probe. The inset on the top right shows the lumped elements of the tip-sample admittance. (**B**) Effective in-plane dielectric constant of monolayer $WS_2$ as a function of the photon energy with an average power of 0.38 μW at 80 K.

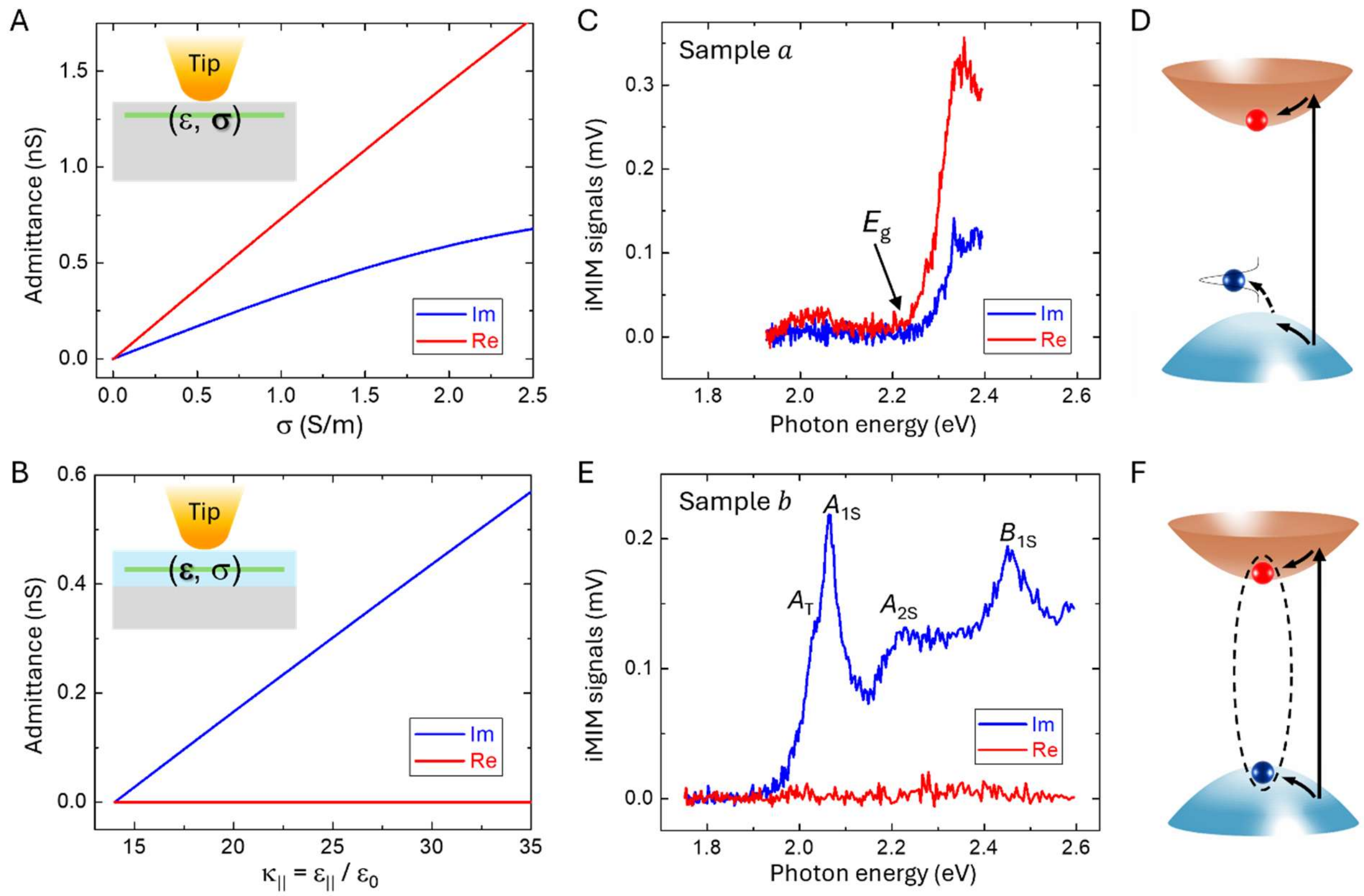

**Fig. 2. Comparison of iMIM response between defect-rich and high-purity samples.** (**A**) Simulated tip admittance as increasing conductivity σ and (**B**) in-plane dielectric constant $\kappa_{\parallel}$ of the 2D layer. The insets show the tip-sample configurations for the FEA simulation. The $WS_2$ and hBN layers are shown as green lines and light blue regions, respectively. (**C**) iMIM signals on the defect-rich Sample $a$. The electronic gap $E_g$ is marked on the plot. (**D**) Schematic of carrier generation, thermalization, and trapping by band-tail states in Sample $a$. (**E**) iMIM signals on the high-purity Sample $b$. The iMIM-Im data exhibits characteristic exciton peaks, whereas the iMIM-Re data remains zero throughout the entire spectrum. (**F**) Schematic of carrier generation, thermalization, and formation of excitons in Sample $b$.

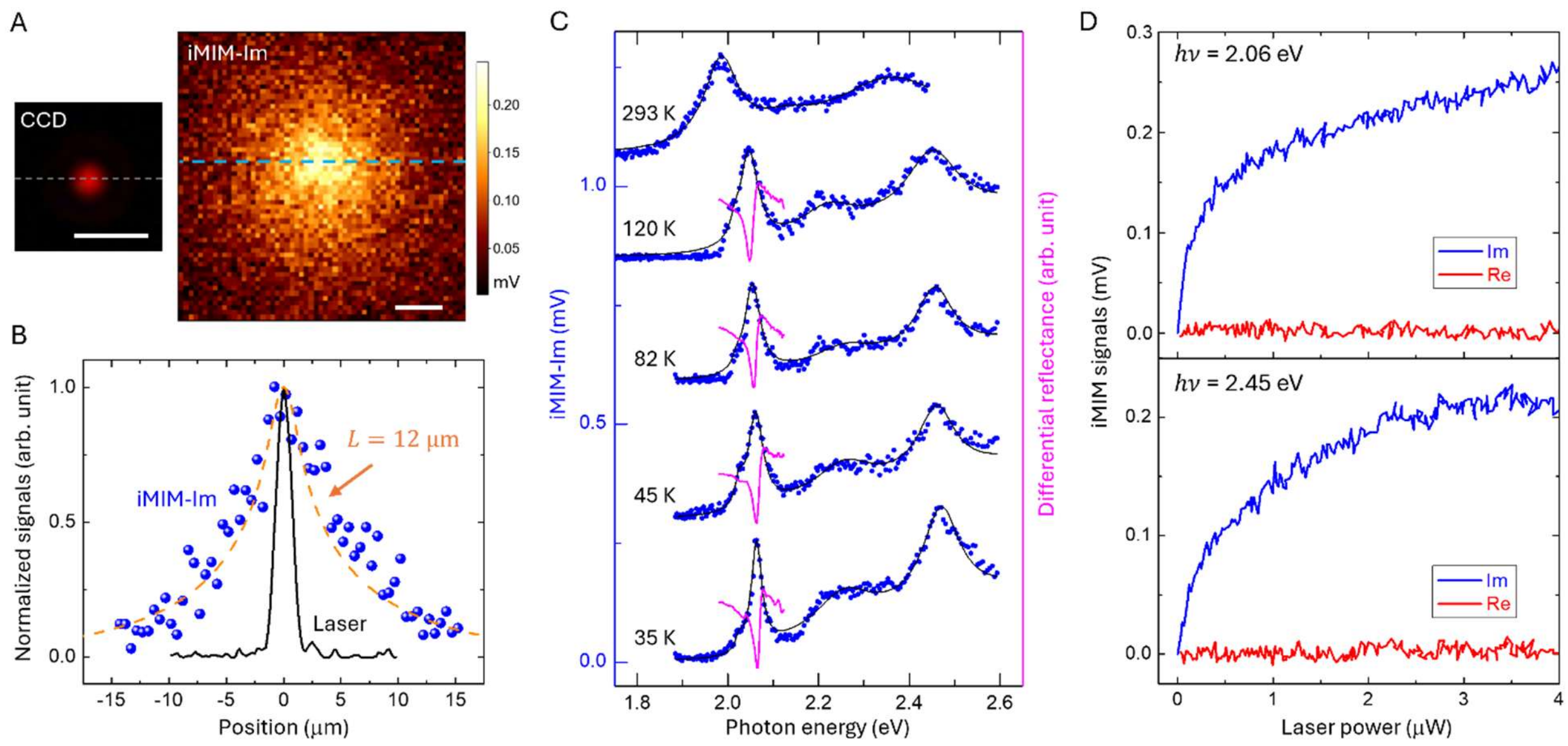

**Fig. 3. Spatial diffusion mapping, temperature and laser-power dependence.** (**A**) CCD image of the laser spot (left) and the iMIM-Im spatial diffusion map (right). The corresponding iMIM-Re map (Supplementary Material S4) is featureless. The scale bars are 5 μm. (**B**) Line profiles across the CCD and iMIM-Im images. The dashed line shows the diffusion profile with a diffusion length $L$ = 12 μm. (**C**) $T$-dependent iMIM-Im data overlaid with optical differential reflectance curves. The solid lines are multi-Lorentzian fits to the iMIM data. The curves are offset vertically for clarity. The corresponding iMIM-Re signals (Supplementary Material S7) are zero under all temperatures. The slight mismatch between $A_{1S}$ peaks in the iMIM data and inflection points in the reflectance curves could be due to different flakes in the measurements (Supplementary Material S8). (**D**) Laser-power dependence of iMIM signals when the photon energy matches the A-exciton (upper) and B-exciton (lower) resonances of monolayer $WS_2$.

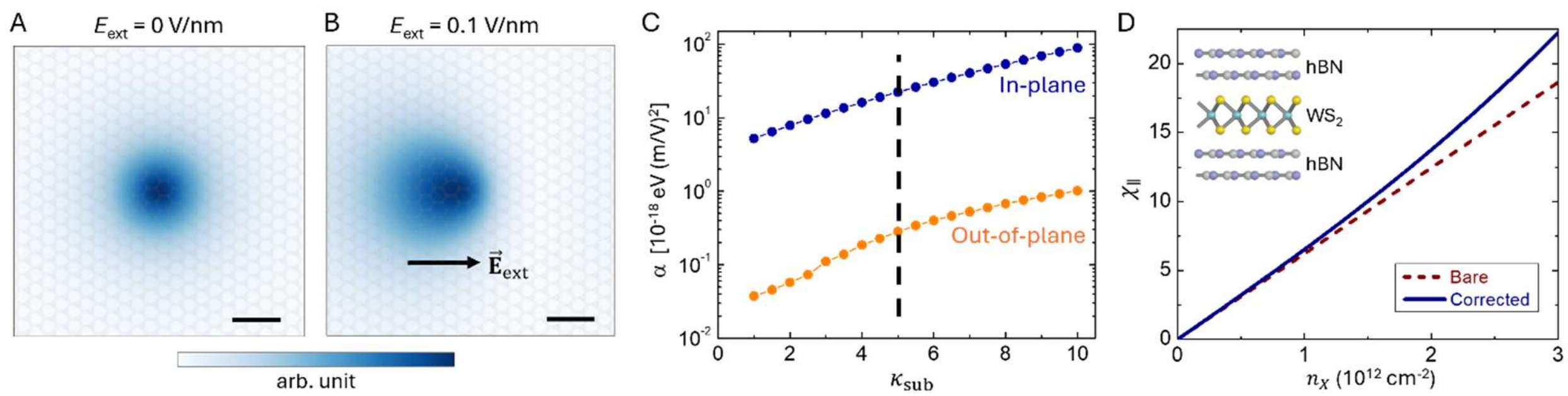

**Fig. 4. Theoretical analysis of the exciton susceptibility in hBN-encapsulated monolayer $WS_2$.** (**A**) Exciton wavefunction in the absence of an external electric field and (**B**) under an in-plane field of 0.1 V/nm. The $WS_2$ lattice is overlaid as the background in the images. The scale bars correspond to 1 nm. (**C**) Calculated in-plane and out-of-plane polarizability values as a function of the effective isotropic dielectric constant of the substrate. The dashed line corresponds to the value of hBN. (**D**) Calculated bare and corrected exciton susceptibility as a function of the exciton density. The inset shows the sample configuration with monolayer $WS_2$ embedded in hBN.